\documentclass[journal]{IEEEtran}
\usepackage{cite}
\usepackage{color}
\usepackage{soul}
\usepackage{epsf}
\usepackage{epsfig}
\usepackage{graphicx}
\usepackage{graphics}
\usepackage{epstopdf}
\usepackage[footnotesize]{caption}
\usepackage{amsmath}
\usepackage{amssymb}
\usepackage{amsthm}
\usepackage{amsxtra}
\usepackage[ruled]{algorithm2e}
\usepackage{algorithmic}
\usepackage{enumerate}
\usepackage{multirow}
\usepackage{enumerate}
\usepackage{amssymb}
\usepackage{lipsum}
\usepackage{setspace}
\usepackage{flushend}
\usepackage{multirow}
\usepackage{subcaption}
\usepackage{geometry}
\usepackage{fancyhdr}
\newtheorem{remark}{Remark}

\usepackage{rotating} 
\usepackage{graphicx} 

\newcommand{\Rmnum}[1]{\expandafter\@slowromancap\romannumeral #1@}

\makeatother
\begin{document}
\title{Management of Renewable Energy for A Shared Facility Controller in Smart Grid}
\author{Wayes~Tushar,~\IEEEmembership{Member,~IEEE,}~Jian~Andrew~Zhang,~\IEEEmembership{Senior Member,~IEEE,}~Chau~Yuen,~\IEEEmembership{Senior Member,~IEEE,}~David~Smith,~\IEEEmembership{Member,~IEEE}~and~Naveed Ul Hassan,~\IEEEmembership{Member,~IEEE}
\thanks{W. Tushar and C. Yuen are with the Singapore University of Technology and Design (SUTD), 8 Somapah Road, Singapore 487372 (Email: \{wayes\_tushar, yuenchau\}@sutd.edu.sg).}
\thanks{J. A. Zhang is with the University of Technology Sydney, Global Big Data Technologies Centre,  Ultimo NSW 2007, Australia (Email: andrew.zhang-1@uts.edu.au).}
\thanks{D. Smith is with Data61 (NICTA), CSIRO, and an Adjunct Fellow with Australian National University (ANU), Australia (Email: david.smith@data61.csiro.au).}
\thanks{N. U. Hassan is with the Lahore University of Management Sciences, Lahore, Pakistan (Email: naveed.hassan@lums.edu.pk).}
\thanks{This work is supported in part by the Singapore University of Technology and Design (SUTD) under grant NRF2015ENC-GBICRD001-028 and NRF2012EWT-EIRP002-045, in part by the Australian Government through the Department of Communications and the Australian Research Council, and in part by the SUTD-MIT International Design Center (IDC) under grant IDG31500106. Any findings, conclusions, recommendations, or opinions expressed in this document are those of the authors and do not necessary reflect the views of the IDC.}}

\IEEEoverridecommandlockouts
\maketitle
\thispagestyle{fancy}
\begin{abstract}
This paper proposes an energy management scheme to maximize the use of solar energy in the smart grid. In this context, a shared facility controller (SFC) with a number of solar photovoltaic (PV) panels in a smart community is considered that has the capability to schedule the generated energy for  consumption and trade to other entities. Particularly, a mechanism is designed for the SFC to decide on the energy surplus, if there is any, that it can use to charge its battery and sell to the households and the grid based on the offered prices. In this regard, a hierarchical energy management scheme is proposed with a view to reduce the total operational cost to the SFC. The concept of a virtual cost (VC) is introduced that aids the SFC to estimate its future operational cost based on some available current information. The energy management is conducted for three different cases and the optimal cost to the SFC is determined for each case via the theory of maxima and minima. A real-time algorithm is proposed to reach the optimal cost for all cases and some numerical examples are provided to demonstrate the beneficial properties of the proposed scheme.
\end{abstract}
 \setcounter{page}{1}
\section{Introduction}\label{sec:introduction}
\IEEEPARstart{T}{he} smart grid provides a suitable platform to accommodate renewable energy sources (RESs) that can provide users with clean energy and thus alleviate users' dependence on conventional power plants~\cite{Fang-J-CST:2012}.  Therefore, the consumers can enjoy green energy for their day to day usage as well as considerably curtailing their energy costs by reducing their reliance on expensive electricity from the main grid~\cite{Tushar-TIE:2014,Tushar-TSG:2016}.

As a consequence, a large number of studies have been conducted on how to accommodate RESs in smart grid through different energy management scheme. These studes can be divided into two general categories. The first category discusses the management aspect via predicting the generation from RESs~\cite{Kalogirou:2014}; and controlling the electricity generation and consumption by demand response management in microgrids through decentralized, distributed and hierarchical control mechanisms~\cite{Guerrero-TIE:2011,Hill-JTSG:2012}. Further, a number of study has explored various energy scheduling schemes for RESs by studying operational management and planning of smart buildings~\cite{Zhang-J_ECM:2013,Kanchev-TIE:2011}, optimization of integrated PV solar houses~\cite{Matrawy:2015}, and efficient building management via distributed predictive control~\cite{Scherer:2014}.

The second category of studies, on the other hand, focuses on energy management techniques for residential smart grid. For instance, in \cite{Muratori-TPS-2016}, a dynamic energy management framework is used to simulate automated residential demand response based on energy consumption models. The models estimate the residential demand that quantifies consumer energy usage behavior and provide an accurate estimation of the controllable resources. A system-wide demand response management model to coordinate demand response provided by residential customers is presented in \cite{Safdarian-TII-2014} to flatten the total load profile for minimizing the individual cost to the customers,. The authors in \cite{PYi-TSG-2013} propose an opportunistic scheduling scheme based on the optimal stopping rule as a real-time distributed scheduling algorithm for smart appliances' automation control. The aim is to balance electricity bill reduction and inconvenience resulting from the operation delay. A two level differential game approach is used to implement a demand response scheme for residential users in \cite{Forouzandehmehr-TSG-2015}. Other energy management schemes for residential smart grid can also be found in \cite{Rassaei-TSE-2015,Yoon-TSG-2014,Tushar-TSG:20162} and \cite{Vivekananthan-TSG-2014}.

\begin{figure}[t]
\centering
\includegraphics[width=\columnwidth]{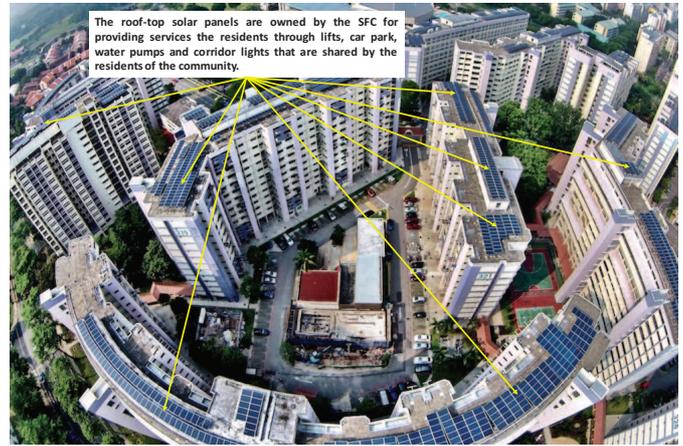}
\caption{Demonstration of the roof-top solar PV panels in a smart community that are owned by the SFC for maintaining facilities that are shared by the residents of the community~\cite{SolarImage:2015}.}
\label{fig:SFCSolar}
\end{figure}

\subsection{Motivation}As can be seen from the above discussion, there has been moderate focus on how to increase the use of RESs for residential smart grid to exclusively use the energy for meeting the entire demand of an entity. Further, most existing techniques on residential energy management have assumed scenarios that involve only two parties: the grid and users with RESs, in which the users decide on either how to schedule their household activities or how to charge and/or discharge their batteries with a view to reduce the respective cost of (or, maximise the benefit of) using electricity from RESs and the grid. Nevertheless, it may also be possible that an entity with RESs may want to exclusively rely on its generated energy for meeting its demand~\cite{Total_Renewable:2015} and  participate in energy trading with multiple heterogeneous energy entities if there is any surplus. This is particularly true for big cities (e.g., Singapore, New York, Shanghai), in which the land is very limited and the population density is very high. Hence, most of the residential accommodations are high rise buildings where it is hard for each unit to install RESs. So the RESs such as solar PVs can only be shared and installed on the building rooftop or car park, and generally these areas are not owned by individuals but rather are controlled by the shared facility controller (SFC) as shown in Fig.~\ref{fig:SFCSolar}. Therefore, unlike the traditional users and grid scenario, such system now involves three parties including the SFC, users and the grid, which has not yet received considerable attention in the literature.

\subsection{Contribution}In this context, this paper proposes a hierarchical renewable energy management scheme that aids the decision making process of an SFC on its energy trading process with multiple energy entities in smart grid. In particular, we study how an SFC can rely exclusively on its generated energy for meeting its own demand (if possible) and then participate in trading with other entities if there is any surplus. Now, due to random changes in energy demand by the users and because the SFC may also want to store some of its generated energy in its energy storage devices (ESDs) for future use,  the SFC needs to decide on 1)~how much surplus energy needs to charge the ESD; 2)~how much surplus energy needs to sell to the households and 3) how much surplus energy needs to sell to the grid such that its can reduce its operational cost of running the shared facilities at any given time.

In this regard, we prioritize the demand of the SFC and allow the SFC to fulfil its demand through its own generation. This is mainly because the SFC owns the solar PVs, as we will see later, and also has its own energy demand for maintaining the shared facilities. Hence, it is reasonable to assume that the SFC will first fulfil its own demand and then use its surplus energy, if there is any, for trading with other energy entities within the system. To leverage the energy management, we introduce the idea of a virtual cost (VC), which is essentially the estimate of a future cost to the SFC based on some information available at current and previous time slots. We use the VC to determine the amount of energy that the SFC can charge to or discharge from its ESD and then use this information to determine how much amount of energy the SFC can trade with the households and the grid in each considered time slot.

Please note that the VC, as we will see in detail in Section~\ref{sec:virtual_Cost}, is important in determining the cost to the SFC. This is due to the fact that a very high VC can let the SFC to store a very large amount of energy in its storage for using in the future, which could be inefficient if the demand is not very high in the future time slot. On the other hand, a very low VC can make the SFC to decide on an amount of energy that could be insignificant compared to the actual requirement in the future. Therefore, the decision on VC needs to be adaptive over time and should possess a realistic value that is comparable to the actual cost incurred to the SFC. We categorise the management problem into three cases based on intermittent solar generation, unpredictable SFC's demand and household demand, and obtain the optimal solution for each of the three cases using the theory of maxima and minima. Furthermore, to coordinate the energy management in real-time, we propose an algorithm that can determine the optimal amount of energy to charge or discharge the ESD of the SFC, so as to attain the optimal operational cost under given system constraints of each case.
\begin{table*}[t]
\centering
\caption{Demonstration of the differences between \cite{Tushar-TIE:2014} and the proposed work.}
\includegraphics[width=\textwidth]{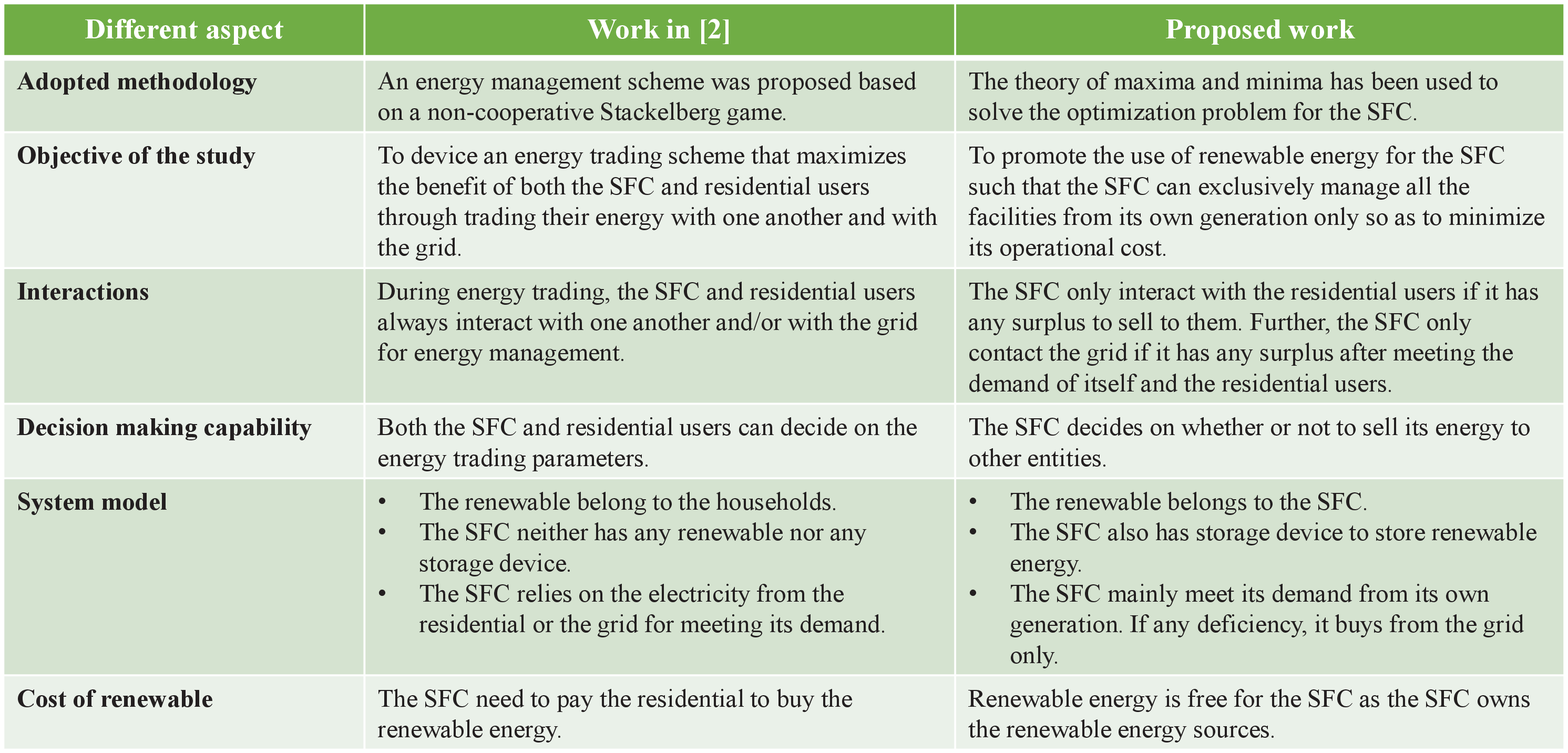}
\label{Fig:table}
\end{table*}

\subsection{Difference with the existing studies}To this end, the main differences between the existing schemes and the proposed work can be summarized as follows:
\begin{itemize}
\item In most energy management studies, the exchange or trading of electricity is designed between two entities, In contrast, we have designed the energy management scheme for the SFC in such a way that it can trade its excess energy, if there is any, with multiple other energy entities within the system. In particular, the proposed scheme involves three parties including the SFC, users and the grid, which has not yet received considerable attention in managing renewable energy. Furthermore, we consider an additional storage device that also forms part of the SFC's decision making variable.
\item In \cite{Tushar-TIE:2014}, a three-party energy model was proposed using a system model that contains the similar components to the proposed study. However, the contents of this paper are substantially different from \cite{Tushar-TIE:2014} as shown in Table~\ref{Fig:table}. Furthermore, this study mainly focus on the interest of the SFC in the smart community whereas in \cite{Tushar-TIE:2014} the authors studied the usefulness of distributed energy resources, which belong to the households, in smart grid.
\item In terms of reducing the cost of energy trading, most of the existing studies focus on the price per unit of energy at different times of the day. That is, the owner of renewable energy sources sells its energy (either from the sources or from the battery) to others when the price is high and buys energy (or stores it) when the price is lower. On the contrary, we focus on promoting the use of renewable energy as much as possible for the SFC and then plan the trading of energy in such a way that the operational cost to the SFC is minimized. To do so, we propose a novel idea of VC, which is a combination of predicted price and predicted demand, to perform energy management.
\end{itemize}

\subsection{Organization of the paper}The remainder of the paper is organised as follows. We discuss the considered system model and problem formulation in Section~\ref{sec:section-2}. The proposed energy management scheme and the algorithm to reach the optimal solution are studied in Section~\ref{sec:section-3}. We analyse the properties of the scheme through numerical case studies in Section~\ref{sec:section-4}, and finally some concluding remarks are drawn in Section~\ref{sec:conclusion}.

\section{System Model and Problem Formulation}
\label{sec:section-2}
\begin{figure}[t!]
\centering
\includegraphics[width=\columnwidth]{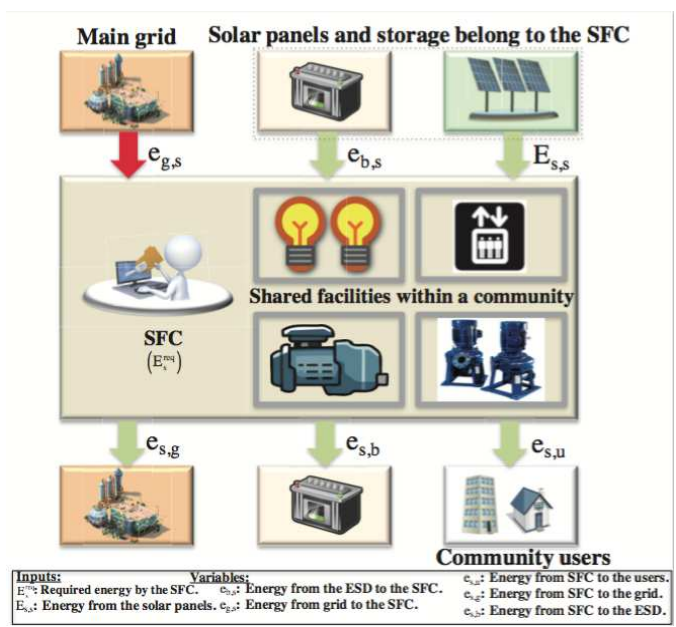}
\caption{The system model for renewable energy management of an SFC, which is is connected to a number of other entities in the smart community.}
\label{fig:SystemModel}
\end{figure}
As in Fig.~\ref{fig:SystemModel}, let us consider a smart community consisting of an SFC, a large number of households and the grid. The SFC is equipped with $N_s$ solar PV panels and an ESD with capacity $B_c$. The area and efficiency of each solar PV panel are assumed to be $A_s$ square meter ($\text{m}^2$) and $\eta$ respectively. Now, if the solar irradiance is $I_\text{pv}(t)$ W/m$^2$ at any particular time $t\in T$, where $T$ is the total considered time duration, the generated power $E_{s,s}(t)$ from the SFC's solar PV is\footnote{Please note that in \eqref{eqn:solar-generation} we use a simple relationship to capture the electricity generation from solar radiation. However, although \eqref{eqn:solar-generation} is simple, we will see in Section \ref{sec:section-3} that the proposed management technique is capable to capture any further variation in the solar power generation due to change in other parameters such as temperature as well.}~\cite{PV-basic,Wayes_ITS:2016}
\begin{eqnarray}
E_{s,s}(t) = \eta\times A_s\times N_s\times I_\text{pv}(t).
\label{eqn:solar-generation}
\end{eqnarray}
At time slot $t$, let us assume that the energy required by the SFC is $E_s^\text{req}(t)$ for the maintenance of different shared facilities of the community and the SFC uses $E_{s,s}(t)$ to meet this requirement. However if $E_{s,s}(t)>E_s^\text{req}(t)$, the SFC may sell the excess energy $\left(E_{s,s}(t) - E_s^\text{req}(t)\right)$ either to the households of the community or to the grid so as to make some extra revenue. Alternatively, the SFC may choose to store this surplus in its ESD for use at a later time. Nonetheless if $E_{s,s}(t)<E_s^\text{req}(t)$, the SFC  either buys the deficient amount of energy $\left(E_s^\text{req}(t)-E_{s,s}(t)\right)$ from the grid or either partially, or completely, discharges its ESD to compensate for this deficiency. We assume that an associated cost\footnote{Please note that revenue can equivalently be considered to be a negative cost.} is always incurred by the SFC whenever the SFC trades energy with the grid and households, and charges or discharges its ESD. Please note that here we use two different variables $e_{b,s}(t)$ and $e_{s,b}(t)$ to refer to the discharging energy from and charging energy to the ESD respectively. These two variables are mainly used to facilitate the decision making process of the SFC on its energy management based on different proposed cases as we will see shortly in Section~\ref{sec:section-3}.
\subsection{Real Cost to the SFC}
\label{subsection:cost function of SFC}
At any time $t$, the total cost $J(t)$ to the SFC is assumed to consist of the following four elements:
\begin{enumerate}
\item Cost $J_\text{buy}(t)$ of buying energy: It is assumed that the SFC buys its deficient energy, if there is any, only from the grid. Hence, the cost $J_\text{buy}(t)$ to the SFC can be expressed as
\begin{eqnarray}
J_\text{buy}(t) = p_{g,s}(t)e_{g,s}(t),
\label{eqn:cost-buy-1}
\end{eqnarray}
where $e_{g,s}(t)$ is the amount of energy that the SFC buys from the grid and $p_{g,s}(t)$ is the price per unit of $e_{g,s}(t)$. It is assumed that the SFC can also discharge $e_{b,s}(t)$ amount of energy from its ESD, if it requires, which may affect the amount of energy it buys from the grid, and the associated costs. Hence, \eqref{eqn:cost-buy-1} can be defined as
\begin{eqnarray}
J_\text{buy}(t) = p_{g,s}(t)([E_s^\text{req}(t) \nonumber\\- \left(E_{s,s}(t) + e_{b,s}(t)\right)]^{+}),
\label{eqn:cost-buy-2}
\end{eqnarray}
where $\left[*\right]^+ = \max\left(0,*\right)$.
\item Revenue $J_\text{sell}(t)$ from selling energy: The revenue of the SFC from selling its surplus energy, if there is any, stems from the revenue $J_\text{user}(t)$ from selling energy $e_{s,u}(t)$ to the household users, and the revenue $J_\text{grid}(t)$ from selling $e_{s,g}(t)$ to the grid. Therefore, the revenue of the SFC from selling its excess generated energy can be expressed as
\begin{eqnarray}
J_\text{sell}(t) =J_\text{user}(t) + J_\text{grid}(t)= -p_{s}(t)\left(e_{s,u}(t)\right) -\nonumber\\ p_{g,\text{buy}}(t)\left(e_{s,g}(t)\right).
\label{eqn:rev-sell-1}
\end{eqnarray}
Here, $p_s(t)$ and $p_{g,\text{buy}}(t)$ are the selling price per unit energy to the users and the grid respectively, and the negative sign implies revenue (instead of cost) for the SFC. Now, to design the revenue $J_\text{sell}(t)$, two factors are considered:
\begin{itemize}
\item The price per unit of energy sold to the users is considerably higher than the selling price to the grid~\cite{Tushar-TIE:2014}. Therefore, it is reasonable to assume that the SFC would want to sell the surplus to the households first, and then, if there is any left over, sell to the grid.
\item The SFC charges its ESD with $e_{s,b}(t)$  from the \emph{excess solar} energy, if there is any.
\end{itemize}
Incorporating these two factors, $J_\text{users}(t)$ and $J_\text{grid}(t)$ in \eqref{eqn:rev-sell-1} are further defined as
\begin{eqnarray}
J_\text{user}(t) = -p_s(t)\min([E_{s,s}(t) -\nonumber\\ (E_s^\text{req}(t)+ e_{s,b}(t))]^+, E_u^\text{req}(t)),
\label{eqn:rev-sell-2}
\end{eqnarray}
where $E_u^\text{req}(t)$ is the energy required by the household users at $t$, and
\begin{eqnarray}
J_\text{grid}(t) = -p_{g,\text{buy}}(t)[E_{s,s}(t) - E_s^\text{req}(t) -\nonumber\\ (E_u^\text{req}(t) + e_{s,b}(t))]^+
\label{eqn:rev-sell-3}
\end{eqnarray}
respectively. From \eqref{eqn:rev-sell-3}, we note that the SFC only sells energy back to the grid when it has enough surplus after meeting its own requirement and the requirements of the household users.
\item Cost $J_\text{SD}(t)$ of charging and discharging of ESD: To limit the number of charging and discharging cycles of the ESD, we consider a cost associated with every charging and discharging of the SFC's ESD~\cite{Deng-eEnergy-2013}, which is assumed to be
\begin{eqnarray}
J_\text{SD}(t) = \alpha_b\max(e_{b,s}(t),e_{s,b}(t)),
\label{eqn:rev-sell-4}
\end{eqnarray}
where $\alpha_b>0$ is the characteristic constant of the ESD, which, for example, can be a function of new ESD cost, depth of discharge, and total number of charging and discharging cycles~\cite{Deng-eEnergy-2013}.
\end{enumerate}

\begin{table*}[t]
\centering
\caption{A toy example to demonstrate the effectiveness of the proposed multi-entity energy trading scheme in reducing the total cost to the SFC through the exploitation of the introduced VC.}
\includegraphics[width=\textwidth]{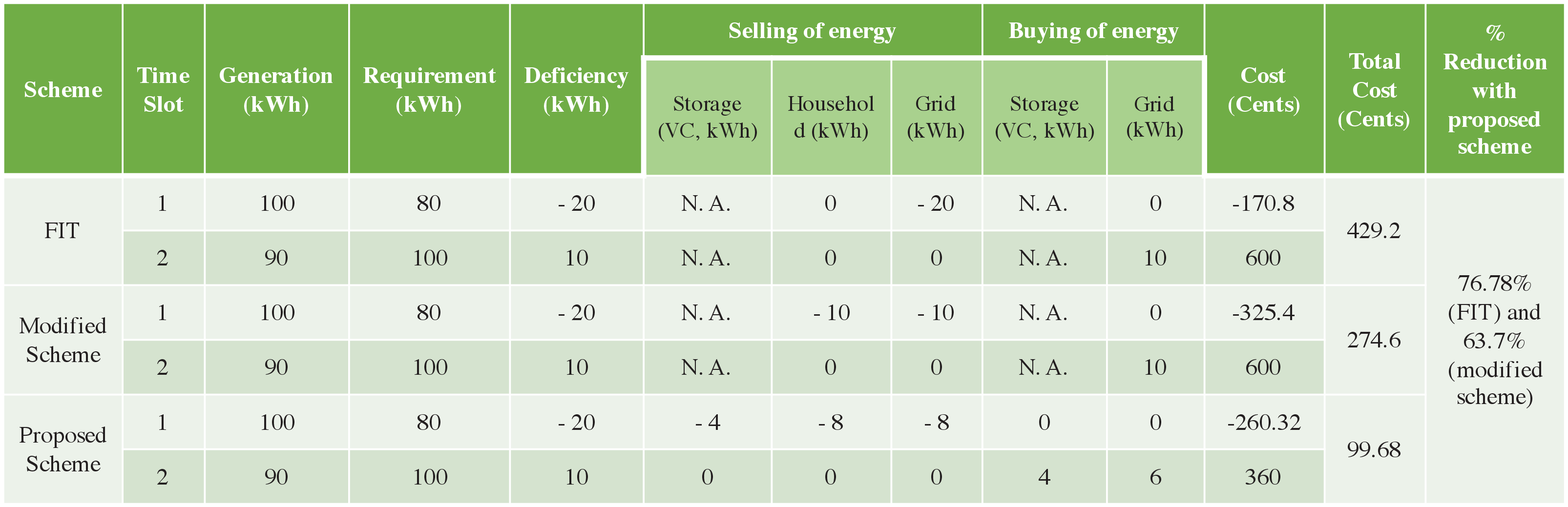}
\label{Table:toyexample}
\end{table*}

\subsection{Virtual Cost}\label{sec:virtual_Cost}
To account for the impact of SFC's current energy decision on the future cost, we introduce a VC $J_v(t)$, which is an estimate of the cost to the SFC at $t+1$ in the current time slot $t$. Essentially, ds{t}he idea of VC is beneficial in assisting the SFC to take the decision on both charging energy to/discharging energy from the storage and on trading with other entities in each time slot of the operation. This can easily be shown by a toy example as follows.

Let us consider a smart community in which an SFC is equipped with solar PV that generates $100$ and $90$ kWh of electricity in time slot $1$ and $2$ respectively, whereby the requirements of the SFC are respectively $80$ and $100$ kWh for the considered two time slots. The selling price per unit of energy to the grid and households are assumed to be $8.54$ and $24$ cents/kWh respectively\footnote{Example of such difference between buying and selling price can be found in Australian Electricity Market, e.g., in the state of Queensland~\cite{Queensland:2015,QueenslandFeed-in-Tariff:2015}.} whereas the buying price from the grid is $60$ cents/kWh~\cite{Tushar-TIE:2014}. For this example, we consider the total cost to the SFC for three techniques: 1) an FIT scheme \cite{McKenna-JIET:2013} where the SFC does not have any capacity to estimate the future cost and sells its excess energy only to the grid; 2) a modified scheme in which the SFC does not have any capacity to estimate the future cost and sells its excess energy to both the grid and the households (such as in \cite{Tushar-TIE:2014}); and 3) a scheme such as the scheme proposed in this paper where the SFC can estimate a relative cost of the next time slot, storing/dispatching energy from its storage accordingly, and can sell the excess energy to both the households and the grid. For all schemes, it is considered that the SFC can buy energy only from the grid. Now, the associated total costs to the SFC for all three schemes are shown in Table~\ref{Table:toyexample}. Please note that the VC is only considered for time slot $1$ as the example assumes only two time slots of energy management. As can be seen for Table~\ref{Table:toyexample}, the proposed scheme has the capability to reduce the total cost to the SFC by $76.7\%$ and $63.7\%$ compared to the FIT and the modified scheme respectively, and thus demonstrates the potential of the proposed scheme in managing renewable energy for the SFC.

We stress that the estimation of future information based on historic values has been widely discussed in the literature. For example, Markov models \cite{Mathieu:2013} and Kalman filtering \cite{Rigatos:2012} have been used extensively for estimating future states based on the historical data of electrical system.  The weighted least square estimator is commonly used in today's power systems, which are sensed primarily via SCADA measurements \cite{GOL:2014}. In \cite{ALAM:2014}, the authors exploits the principles of one-dimensional and two-dimensional compressive sensing to develop approaches for voltage estimation in distribution grids with renewable energy based generators. Other techniques related to estimation of states in dynamical systems can also be found in \cite{Powell:2007}. However, in contrast to these techniques, we adopt a simple technique to determine the virtual cost, which is suitable for the considered system model. In particular, the proposed virtual cost model depends only on the information available at the current and previous time slots and therefore does not require memory to store a large amount of historical data. Thus, the virtual cost is computationally less expensive than existing techniques, and simulation results validate its effectiveness, as will be shown in Section~\ref{sec:section-4}.

To this end, the proposed $J_v(t)$ accounts for the impact of the state-of-charge (SoC) of the SFC's ESD on its future cost. For instance, if the SoC of the ESD is small at the end of optimization at time slot $t$, the cost to the SFC could be estimated to be larger at $t+1$ as the amount of energy $e_{b,s}(t+1)$ that could be discharged from the ESD would be small. Hence, the SFC may be required to buy more energy from the grid if the generation of solar energy $E_{s,s}(t+1)$ is not enough to meet its requirement $E_s^\text{req}(t+1)$. In this context, we assume that $J_v(t)$ is a decreasing function of the SoC $s_b(t)$ at the end of time slot $t$. Mathematically,
\begin{eqnarray}
J_v(t) &=& \frac{a(t)}{s_b(t)}\nonumber\\&=&\frac{a(t)}{s_b(t-1) + \nu(e_{s,b}(t) - e_{b,s}(t))}.
\label{eqn:virtual-cost-1}
\end{eqnarray}
Here, $\nu$ is the efficiency of the SFC's ESD and $a(t)>0$ is a coefficient, which is adaptive across time slots. The motivation for using an adaptive $a(t)$ is to better capture the  non-linear effect of current energy flow on the virtual cost. For example, if  the SFC is buying more energy from the grid in the current time slot, it may need to buy more energy in the next time slot too, due to the time coherence in energy consumption. Hence the updating function of $a(t)$ is proposed to follow the relation
\begin{eqnarray}
a(t) = a(t-1) + \mu\left(e_{g,s}(t-1) - e_{g,s}(t-2)\right),
\label{eqn:update-function}
\end{eqnarray}
where $\mu>0$ is a scalar step parameter. It is important to note that the value $a(t)$ is updated based on the amount of $e_{g,s}$ that the SFC buys from the grid. The initial value $a_\text{ini}$ of $a(t)$ is considered to be a design parameter, which depends on various system parameters such as community size. Hence, if the size of the community is larger and/or number of solar PVs installed in the SFC is smaller, the SFC will need to buy more energy from the grid, which will consequently increase the value of $a_\text{ini}$ and the VC. Therefore, the choice of $a_\text{ini}$ is significantly affected by the type and size of the community and the capacity of the installed solar PVs. Furthermore, if the value of $\mu$ becomes higher, the value of $a(t)$ and, consequently, the value of $J_v(t)$ increases for purchasing a similar amount of energy from the grid. To this end, $\mu$ can be referred to as the sensitivity of SFC's cost to the energy that it purchases from the grid. As we will see shortly, the considered energy management scheme considerably depends on the proposed VC model.

\subsection{Total Cost and Objective Function}
Combining all the relevant costs from \eqref{eqn:cost-buy-1}, \eqref{eqn:rev-sell-1}, \eqref{eqn:rev-sell-4} and \eqref{eqn:virtual-cost-1}, the cost function of the SFC can be defined as
\begin{eqnarray}
J(t) = J_\text{buy}(t) + J_\text{sell}(t) + J_\text{SD}(t) + J_v(t),
\label{eqn:cost-function-1}
\end{eqnarray}
for each $t = 1, 2, \hdots, T$. It is important to note that one of the objectives of this work is to maximize the use of generated solar energy in order to minimize the cost to the SFC. Therefore, the problem is formulated such that the SFC only buys energy from the grid when the generated solar  $E_{s,s}(t)$ is not large enough to meet its requirement $E_s^\text{req}(t)$. Also, the SFC only sells its energy when $E_{s,s}(t)>E_s^\text{req}(t)$. Furthermore, it can be seen from \eqref{eqn:cost-buy-2}, \eqref{eqn:rev-sell-2}, \eqref{eqn:rev-sell-3}, \eqref{eqn:rev-sell-4} and \eqref{eqn:virtual-cost-1} that all the related costs to the SFC can be expressed in terms of its charging and discharging amount of energy $e_{b,s}(t)$ and $e_{s,b}(t)$ respectively. To that end, the objective of the SFC can be expressed in terms of minimizing $J(t)$ by choosing a suitable either $e_{b,s}(t)$ or $e_{s,b}(t)$ in each time slot $t$, which can be defined mathematically as
\begin{eqnarray}
\min_{e_{b,s}(t)~\text{or}~e_{s,b}(t)}\left[J_\text{buy}(t) + J_\text{sell}(t) + J_\text{SD}(t) + J_v(t)\right],
\label{eqn:cost-function-2}
\end{eqnarray}
for $t = 1, 2, \hdots, T$. Now, to solve the problem \eqref{eqn:cost-function-2} in real time, we propose an energy management scheme in the next section.
\section{Energy Management Scheme}
\label{sec:section-3}
The energy management process of the SFC at any time slot $t$ always falls within one of the following three categories based on intermittent solar generation and random demand of the SFC and the household users:
\begin{enumerate}
\item Case 1: $E_s^\text{req}(t)\geq E_{s,s}(t)$. The SFC does not sell any energy. The SFC may buy electricity from the grid and/or discharge its ESD.
\item Case 2: $E_s^\text{req}(t)<E_{s,s}(t) \leq \left(E_s^\text{req}(t) + E_u^\text{req}(t)\right)$. The SFC does not buy any electricity from the grid and may charge its ESD. The SFC may sell electricity to the household users but not to the grid.
\item Case 3: $E_{s,s}(t) > \left(E_s^\text{req}(t) + E_u^\text{req}(t)\right)$. The SFC does not buy any electricity and may charge its ESD. The SFC may sell electricity to the household users and the grid.
\end{enumerate}
Now, what follows is a detailed analysis of the optimal cost to the SFC, in each of the three cases, through the derivation of the energy amount that the SFC needs to either charge or discharge to reach the optimal cost.
\subsection{Case 1}
\label{subsection:3.1}
In this case, the SFC neither sells any electricity nor charges its ESD. Therefore, $e_{s,u}(t), e_{s,g}(t), e_{s,b}(t)=0$. So, the cost function $J_\text{case-1}(t)$ for case $1$ reduces to the form\footnote{Hereinafter, the cost function of the SFC for each case $i$,~where $i = 1,2,3$, will be indicated as $J_\text{case-i}$.}
\begin{eqnarray}
J_\text{case-1} (t) = p_{g,s}(t)\left(E_s^\text{req}(t) - (E_{s,s}(t) + e_{b,s}(t))\right) +\nonumber\\ \alpha_b e_{b,s}(t)+ \frac{a(t)}{s_b(t-1) - \nu e_{b,s}(t)},
\label{eqn:cost-case-1}
\end{eqnarray}
where the objective of the SFC is
\begin{eqnarray}
\min_{e_{b,s}(t)} J_\text{case-1} (t).
\label{eqn:obj-case-1}
\end{eqnarray}
Now, $J_\text{case-1}(t)$ attains its minimum value\footnote{Since, $\frac{\delta^2 J_\text{case-1}(t)}{\delta e_{b,s}(t)^2}=\frac{2\nu^2 a(t)}{[s_b(t-1) - \nu e_{b,s}(t)]^3}>0.$} for the choice of $e_{b,s}(t)$ when $\frac{\delta J_\text{case-1}(t)}{\delta e_{b,s}(t)} = 0$, and thus
\begin{eqnarray}
e_{b,s}(t) = \frac{1}{\nu}\left[s_b(t-1) - \sqrt{\frac{\nu a(t)}{p_{g,s}(t) - \alpha_b}}\right].
\label{eqn:optimal-energy-case-1}
\end{eqnarray}
Thus, the optimal cost to the SFC for case $1$ is obtained by \eqref{eqn:cost-case-1} when $e_{b,s}(t)$ is as given in \eqref{eqn:optimal-energy-case-1}.
\subsection{Case 2}
\label{subsection:3.2}
In case $2$, the SFC neither buys any electricity from nor sells any electricity to the grid. Therefore, $e_{g,s}(t) = 0$ and $e_{s,g}(t) = 0$. Also, the ESD does not discharge, i.e., $e_{b,s}(t) = 0$. Hence, the cost function and the objective of the SFC can be expressed as
\begin{eqnarray}
J_\text{case-2}(t) = -p_s(t) (E_{s,s}(t) - (E_s^\text{req}(t)) + e_{s,b}(t)) +\nonumber\\ \alpha_be_{s,b}(t) + \frac{a(t)}{s_b(t-1) + \nu e_{s,b}(t)},
\label{eqn:cost-case-2}
\end{eqnarray}
and
\begin{eqnarray}
\min_{e_{s,b}(t)} J_\text{case-2}(t)
\label{eqn:obj-case-2}
\end{eqnarray}
respectively. Now, the choice of $e_{s,b}(t)$ for which $J_\text{case-2}(t)$ attains the minimum value\footnote{For same reason in Case-1.} satisfies $\frac{\delta J_\text{case-2}(t)}{\delta e_{s,b}(t)} = 0$. Hence, the optimal choice of $e_{s,b}(t)$ for case $2$ is
\begin{eqnarray}
e_{s,b}(t) = \frac{1}{\nu}\left[\sqrt{\frac{\nu a(t)}{\alpha_b + p_s(t)}}-s_b(t-1)\right].
\label{eqn:optimal-energy-case-2}
\end{eqnarray}
\subsection{Case 3}
\label{subsection:3.3}
In this case, the SFC sells electricity to the grid after meeting its own demand and those of the households. However, it does not buy any electricity from the grid, i.e., $e_{g,s}(t) = 0$. To this end, the cost to the SFC for case $3$ can be expressed as
\begin{eqnarray}
J_\text{case-3}(t) = -p_s(t)E_u^\text{req}(t) - p_{g,\text{buy}}(t)(E_{s,s}(t) - E_s^\text{req}(t) -\nonumber\\ (E_u^\text{req}(t) + e_{s,b}(t))) + \alpha_b e_{s,b}(t) + \frac{a(t)}{s_b(t-1) + \nu e_{s,b}(t)}.
\label{eqn:cost-case-3}
\end{eqnarray}
The first term of \eqref{eqn:cost-case-3} refers to the total revenue that the SFC attains from selling the required energy to the households. Now, similar to the previous two cases in Section \ref{subsection:3.1} and Section \ref{subsection:3.2}, the optimal choice of $e_{s,b}(t)$ in order to minimize the SFC's cost $J_\text{case-3}(t)$ can be obtained as
\begin{eqnarray}
e_{s,b}(t) = \frac{1}{\nu}\left[\sqrt{\frac{\nu a(t)}{\alpha_b + p_{g,\text{buy}}(t)}}-s_b(t-1)\right].
\label{eqn:optimal-energy-case-3}
\end{eqnarray}
\subsection{Constraints}
\label{subsection:3.4}
Now, while the SFC minimizes its cost by suitably choosing $e_{b,s}(t)$ or $e_{s,b}(t)$ according to \eqref{eqn:optimal-energy-case-1}, \eqref{eqn:optimal-energy-case-2} or \eqref{eqn:optimal-energy-case-3}, the SFC needs to maintain a number of constraints for suitable implementation of the approach in a practical environment. Some of these constraints are based on the cases proposed in this paper. In the following, we briefly explain the constraints that are assumed to be satisfied by the SFC during management of its energy.
\begin{enumerate}
\item Equality constraint on energy trading: At time slot $t$, the total supply of energy to the SFC should be equal to the total energy spent by the SFC in the considered time slot. That is
\begin{eqnarray}
E_{s,s}(t) + e_{b,s}(t) + e_{g,s}(t) = E_s^\text{req}(t) +\nonumber\\ E_u^\text{req}(t) + e_{s,b}(t) + e_{s,g}(t).\label{eqn:constraint-1}
\end{eqnarray}
\item Constraint on the SOC: The SoC $s_b(t)$ of the SFC's ESD at time slot $t$ is a function of the charging and discharging energy amount, i.e., $e_{s,b}(t)$ and $e_{b,s}(t)$ respectively, at $t$ and the SoC $s_b(t-1)$ from the previous time slot. This relationship can be expressed as
\begin{eqnarray}
s_b(t) = s_b(t-1) + \nu \left(e_{s,b}(t) - e_{b,s}(t)\right),\label{eqn:constraint-battery-1}
\end{eqnarray}
where $\nu$ is the ESD efficiency. Also, the SoC of the EDS at any time slot $t$ cannot be larger than its capacity $B_\text{cap}$ (i.e., $100\%$ SoC) or lower than a certain minimum amount $B_\text{min}$ in order to prolong the life-time of the ESD. Therefore,
\begin{eqnarray}
B_\text{min}\leq s_b(t)\leq B_\text{cap}.\label{eqn:constraint-battery-2}
\end{eqnarray}
\item Constraint on ESD's charging and discharging: The SFC cannot charge and discharge its ESD simultaneously in any time slot $t$. That is
\begin{eqnarray}
e_{s,b}(t)~\text{is}~ \begin{cases}
\geq 0, & \text{if}~ e_{b,s}(t) = 0\\
= 0, &\text{if}~ e_{b,s}(t) > 0
\end{cases},
\label{eqn:constraint-battery-charge-1}
\end{eqnarray}
and
\begin{eqnarray}
e_{b,s}(t)~\text{is}~ \begin{cases}
\geq 0, & \text{if}~ e_{s,b}(t) = 0\\
= 0, & \text{if}~ e_{s,b}(t) > 0
\end{cases}.
\label{eqn:constraint-battery-charge-2}
\end{eqnarray}
The charging and discharging rate of the ESD cannot be greater than the ESD's rated charging/discharging capacity $e_b^\text{max}$. Also, an SFC cannot charge its battery more than the available space in its ESD. Similarly, the SFC cannot discharge its ESD more than the available SoC. Therefore,
\begin{eqnarray}
e_{s,b}(t) \leq \min(e_b^\text{max},(B_\text{cap}-s_b(t-1))), \label{eqn:constraint-battery-charge-3}
\end{eqnarray}
and
\begin{eqnarray}
e_{b,s}(t) \leq \min(e_b^\text{max}, (s_b(t-1)-B_\text{min})). \label{eqn:constraint-battery-charge-4}
\end{eqnarray}
\item Constraint on grid energy: The SFC does not buy energy from and sell energy to the grid at the same time slot. That is
\begin{eqnarray}
e_{g,s}(t)~\text{is}~ \begin{cases}
\geq 0, & \text{if}~ e_{s,g}(t) = 0\\
= 0, &\text{if}~ e_{s,g}(t) > 0
\end{cases},
\label{eqn:constraint-3}
\end{eqnarray}
and
\begin{eqnarray}
e_{s,g}(t)~\text{is}~ \begin{cases}
\geq 0, & \text{if}~ e_{g,s}(t) = 0\\
= 0, & \text{if}~ e_{g,s}(t) > 0
\end{cases}.
\label{eqn:constraint-4}
\end{eqnarray}
\item Constraint on $\alpha_b$: The choice of $\alpha_b$ may affect the optimal choice of $e_{b,s}(t)$ and $e_{s,b}(t)$ of the SFC through \eqref{eqn:optimal-energy-case-1}, \eqref{eqn:optimal-energy-case-2} and \eqref{eqn:optimal-energy-case-3}. Now to decide how to choose a suitable value of $\alpha_b$, we first note from \eqref{eqn:optimal-energy-case-1} that the SFC will discharge its ESD in case $1$ if
\begin{eqnarray}
\alpha_b<p_{g,s}(t) - \frac{\nu a(t)}{(s_b(t-1))^2}.\label{eqn:constraint-alphab-1}
\end{eqnarray}
And, the charging of the ESD in case $2$ and case $3$ takes place if
\begin{eqnarray}
\alpha_b < \frac{\nu a(t)}{(s_b(t-1))^2} - p_s(t)\label{eqn:constraint-alphab-2}
\end{eqnarray}
and
\begin{eqnarray}
\alpha_b<\frac{\nu a(t)}{(s_b(t-1))^2} - p_{g,\text{buy}}(t)\label{eqn:constraint-alphab-21}
\end{eqnarray}
respectively. Since $p_s(t)>p_{g,\text{buy}}(t)$, it is clear that if \eqref{eqn:constraint-alphab-2} is true the condition in \eqref{eqn:constraint-alphab-21} is also true. Now, the conditions on $\alpha_b$ in \eqref{eqn:constraint-alphab-1} and \eqref{eqn:constraint-alphab-2} are satisfied if $\alpha_b$ is set such that
\begin{eqnarray}
\alpha_b<\frac{p_{g,s}(t) - p_s(t)}{2},~\forall t.\label{eqn:constraint-alphab-3}
\end{eqnarray}
In this context, to satisfy the condition in \eqref{eqn:constraint-alphab-3} at each $t = 1, 2, \hdots, T$, the value of $\alpha_b$ needs to be chosen such that
\begin{eqnarray}
\alpha_b<\min\left(\frac{p_{g,s}(t) - p_s(t)}{2},~\forall t\right).\label{eqn:constraint-alphab-4}
\end{eqnarray}
\end{enumerate}
\subsection{Algorithm}
\begin{algorithm}[t]
\caption{Algorithm for the SFC to reach the optimal solution.}
\label{algorithm:1}
\scriptsize
\begin{algorithmic}[1]
\STATE \textbf{Initialization}: $e_{g,s}(t)$ and $a(t)$ for $t \in\{1,2\}$.
  \FOR{Time slot $t = 3$ to $T$}
  \STATE The Grid announces $p_{g,s}(t), p_{g,\text{buy}}(t)$.
  \STATE \textbf{The SFC}:
  \STATE Sets $i=0$.
  \STATE Sets $p_s(t)$.
    \STATE Determines the SOC $s_b(t)$.
    \STATE Calculates the PV generation $E_{s,s}(t)$.
    \STATE Calculates its requirement $E_s^\text{req}(t)$.
    \STATE Receives users total energy requirement $E_u^\text{req}(t)$.
    \STATE Calculates $a(t) = a(t-1) + \mu\left(e_{g,s}(t-1) - e_{g,s}(t-2)\right)$.
    \IF {$E_s^\text{req}(t)>E_{s,s}(t)$}
    \STATE Set $i=1$.
    \STATE Calculates $e_{b,s}(t)$ following \eqref{eqn:optimal-energy-case-1}.
      \IF {$e_{b,s}(t)>\min(e_b^\text{max}, (s_b(t-1)-B_\text{min}))$}
         \STATE Sets $e_{b,s}(t) = \min(e_b^\text{max}, (s_b(t-1)-B_\text{min}))$.
      \ENDIF
      \IF {$e_{b,s}(t)<0$}
         \STATE Sets $e_{b,s}(t) = 0$.
      \ENDIF
        \STATE Calculates $J_\text{case-1}(t)$ using \eqref{eqn:cost-case-1}.
    \ENDIF
   \IF {$E_{s,s}(t)\geq E_s^\text{req}(t)$ and $E_{s,s}(t)<E_s^\text{req}(t) + E_u^\text{req}(t)$}
   \STATE Sets $i=2$.
   \STATE Calculates $e_{s,b}(t)$ from \eqref{eqn:optimal-energy-case-2}.
    \IF {$e_{s,b}(t)>\min(e_b^\text{max},(B_\text{cap}-s_b(t-1)))$}
         \STATE Sets $e_{s,b}(t) = \min(e_b^\text{max},(B_\text{cap}-s_b(t-1)))$.
      \ENDIF
      \IF {$e_{s,b}(t)<0$}
         \STATE Sets $e_{s,b}(t) = 0$.
      \ENDIF
        \STATE Calculates $J_\text{case-2}(t)$ using \eqref{eqn:cost-case-2}.
   \ENDIF
      \IF {$E_{s,s}(t)>E_s^\text{req}(t) + E_u^\text{req}(t)$}
        \STATE Sets $i=3$.
   \STATE Calculates $e_{s,b}(t)$ from \eqref{eqn:optimal-energy-case-3}.
    \IF {$e_{s,b}(t)>\min(e_b^\text{max},(B_\text{cap}-s_b(t-1)))$}
         \STATE Sets $e_{s,b}(t) = \min(e_b^\text{max},(B_\text{cap}-s_b(t-1)))$.
      \ENDIF
      \IF {$e_{s,b}(t)<0$}
         \STATE Sets $e_{s,b}(t) = 0$.
      \ENDIF
        \STATE Calculates $J_\text{case-3}(t)$ using \eqref{eqn:cost-case-3}.
   \ENDIF
   \STATE Sets $J(t) = J_\text{case-i}(t)$.
  \STATE Determines $s_b(t)$ using \eqref{eqn:constraint-battery-1}.
  \STATE Determines $e_{g,s}(t)$ through \eqref{eqn:constraint-1}.
  \ENDFOR
\end{algorithmic}
\end{algorithm}
After determining the optimal cost to the SFC for the proposed three cases, and defining the related constraints, we now introduce an algorithm, which can be adapted by the SFC to reach the optimal solution in real time. The algorithm is initiated in each time slot $t$ through the announcement of $p_{g,s}(t)$ and $p_{g,\text{buy}} (t)~\forall t$ by the grid, and the setting up of $p_s(t)~\forall t$ by the SFC such that \eqref{eqn:constraint-alphab-4} is satisfied. In each time slot $t$, the SFC gets information on its generated energy $E_{s,s}(t)$, determines its requirement $E_s^\text{req}(t)$, and receives the energy request $E_u^\text{req}(t)$ from the household users. Based on the available information, the SFC determines the category of the energy management scheme. Then, according to the type of category, i.e., case $1$, $2$ or $3$, and the associated constraints in Section~\ref{subsection:3.4}, the SFC obtains the optimal charging and discharging amount according to the discussion in Section~\ref{subsection:3.1}, \ref{subsection:3.2} and  \ref{subsection:3.3}. Subsequently, the optimal costs for all three cases are determined. The detail of the proposed algorithm is shown in Algorithm~\ref{algorithm:1} in a step-by-step fashion.

It is important to note that in each iteration only one of the three cases is executed in the algorithm. Further, the decision making on the cost of the SFC in each case is based on the  expressions that have been derived in \eqref{eqn:cost-case-1}, \eqref{eqn:optimal-energy-case-1} (case-1), \eqref{eqn:cost-case-2}, \eqref{eqn:optimal-energy-case-2} (case-2), and \eqref{eqn:cost-case-3}, \eqref{eqn:optimal-energy-case-3} (case-3)  respectively for the three cases. Hence, the implementation of the algorithm is simple and can be executed with minimal computational complexity.

\begin{remark}
Note that all possible scenarios of the considered energy management scheme are captured through the three cases proposed in Section~\ref{subsection:3.1}, \ref{subsection:3.2} and \ref{subsection:3.3}, and the optimal cost to the SFC in each of the considered cases is determined through Algorithm~\ref{algorithm:1}. To determine the feasible optimal solution, Algorithm \ref{algorithm:1} is leveraged via the proposed concept of VC, and is executed according to the derived expressions in Section~\ref{subsection:3.1}, \ref{subsection:3.2} and \ref{subsection:3.3} where the constraints of Section \ref{subsection:3.4} are also maintained. Hence, Algorithm~\ref{algorithm:1} always reaches the optimal solution of the proposed energy management scheme.
\end{remark}

It is important to note that the proposed optimization problem can be solved by following Algorithm 1 due to the obtained expressions in \eqref{eqn:cost-case-1}, \eqref{eqn:optimal-energy-case-1} (case-1), \eqref{eqn:cost-case-2}, \eqref{eqn:optimal-energy-case-2} (case-2), and \eqref{eqn:cost-case-3}, \eqref{eqn:optimal-energy-case-3} (case-3). Nonetheless, other optimization techniques such as particle swarm optimization and simulated annealing may also be suitable to solve the proposed problem.
\section{Case Study}
\label{sec:section-4}
In this section, we provide some numerical simulation results to show the beneficial properties of the proposed scheme. We demonstrate how the proposed scheme can help the SFC to reduce its average cost over a considered period of time, e.g., a day.

We consider that the SFC owns a solar array, e.g., on the rooftop of the community buildings, consisting of $65$ solar panels and has an ESD of capacity $15$ kWh. Each solar panel has a dimension of $1.926\times 1.014$ m$^2$~\cite{SolarPanel} and an efficiency of $0.30$~\cite{solar-efficiency}. Total time duration is considered to be from $6.00$ am to $8.00$ pm, which consists of $28$ time slots and each time slot is assumed to have a duration of $30$ minutes~\cite{Anderson-W-EIS:2010}. The value of solar irradiance at each time slot is taken from the set of solar data (measured at the campus of Australian National University, Canberra, Australia), which is averaged over a month of data for the considered time slots. The grid's real time sell price $p_{g,s}(t)~\forall t$ per unit of electricity is considered from \cite{PJM:2013}, and the sell price $p_s(t)$ of the SFC and the buy price $p_{g,\text{buy}}(t)$ of the grid are assumed to be $0.6$ and $0.3$ times $p_{g,s}(t)$ respectively\footnote{$p_s(t)$ and $p_{g,\text{buy}}(t)$ are chosen such that the condition $p_{g,\text{buy}}(t)<p_s(t)<p_{g,s}(t)$ is always maintained throughout the energy management scheme~\cite{Tushar-TIE:2014}.}. The requirement of the SFC is calculated based on the demand of community lifts at different time of the day. In particular, we choose the number of trips of the community lifts randomly between $[100, 200]$ times during the peak hours, i,e., $6$ am to $9$ am and $4.30$ pm to $8.00$ pm, and between $[70, 100]$ times for the rest of the time. The energy consumption of the lifts for each trip is assumed to be $0.1$ kilo-watt hour (kWh)~\cite{elevator-energy}. Total demand of the households at different time of the day is considered randomly from the range $[10, 25]$ kWh~\cite{ENergy-usage}. Unless stated otherwise, the values of $\alpha_b$  and $a_\text{ini}$ are assumed to be $\min\left(\frac{p_{g,s}(t) - p_s(t)}{2},~\forall t\right)-1$ and $250$ respectively. It is important to note that all parameter values are particular to this study and may vary for different cases based on circumstances such as weather conditions, number of households in a community, electricity price, time of day and the nation (or state) where it is located.

\begin{figure}[t]
\centering
\includegraphics[width=\columnwidth]{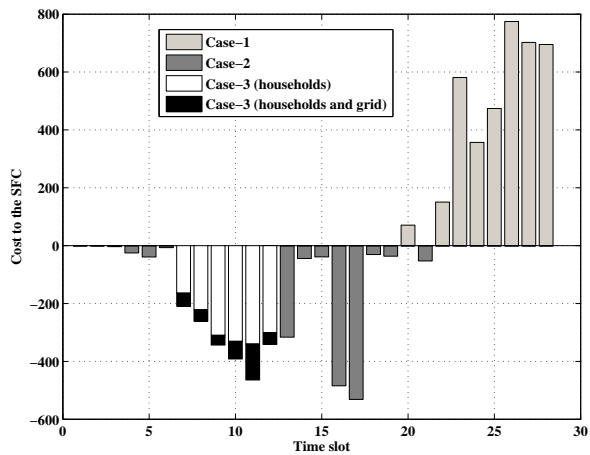}
\caption{Demonstration of cost (in cents) to the SFC at different time slots of the considered time duration. The positive costs are incurred to the SFC during case-1 whereby the negative costs, i.e., the revenue, are attained by the SFC during case-2 and case-3.} \label{fig:variation-of-cost}
\end{figure}
\begin{figure}[t]
\centering
\includegraphics[width=\columnwidth]{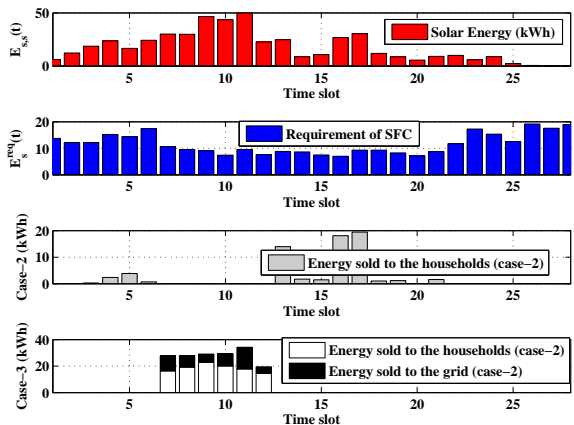}
\caption{Demonstration of the decision making process of the SFC on how much energy it needs to sell to different entities.} \label{fig:energytrading}
\end{figure}
\subsection{Behavior of the scheme at different time slots}
To this end, we first show how the proposed scheme can react to environmental change in the system and execute the energy management scheme for different cases. In particular, we show how the optimization problem of the SFC falls into different cases, i.e., case-1, case-2 and case-3, at different time slots in Fig.~\ref{fig:variation-of-cost} according to the solar generation and SFC's requirements as demonstrated in Fig.~\ref{fig:energytrading}. According to Fig.~\ref{fig:energytrading}, the SFC's solar generation is higher than its required energy at time slots $1$ to $19$ and again at $21$. Therefore, during these time slots, the SFC sells its excess energy to both the households and the grid during time slots $7$ to $12$, and to the households only for the rest of the time. This is due to the fact that when the surplus from the generation is significantly higher, i.e., from time slots $7$ to $12$, the SFC sells to the grid after meeting its own electricity demand and the requirements of the households according to the designed scheme. Thus, these time durations fall within case-3 as shown in Fig.~\ref{fig:variation-of-cost}, bringing revenue for the SFC. For relatively smaller surpluses, i.e., case-2 in Fig.~\ref{fig:variation-of-cost},  the SFC sells its surplus energy only to the households\footnote{Similar to case-3, the SFC also receives revenue in case-2.}, e.g., in time slots $4, 5, 6, 13$ to $19$, and $21$ in Fig.~\ref{fig:energytrading}. Nevertheless, according to Fig.~\ref{fig:energytrading}, at the latter part of the considered range of time slots, the SFC's requirement is significantly higher than the generation. Hence, the SFC needs to buy energy from the grid during these time slots and thus the energy management scheme falls within case-1, as shown in Fig.~\ref{fig:variation-of-cost}. In Fig.~\ref{fig:energytrading}, due to space limitation we do not show the amount of energy that the SFC buys from grid during these time slots, which can easily be calculated from the difference between the solar generation and SFC's requirements. Note that, unlike case-2 and case-3, the energy trading in case-1 incurs cost to the SFC.

\subsection{Change of VC with $a_\text{ini}$}
\begin{figure}[t!]
\centering
\includegraphics[width=\columnwidth]{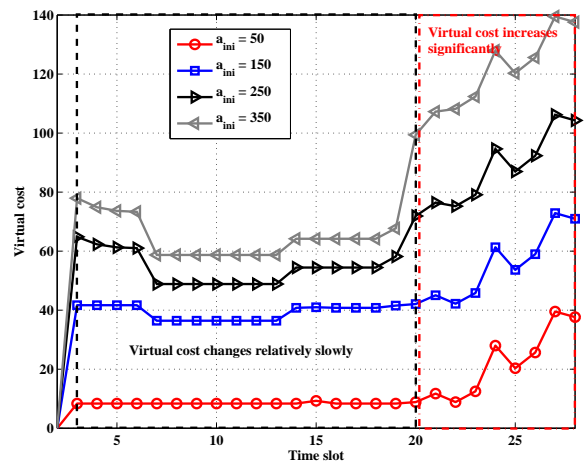}
\caption{Demonstration of the change of VC at different times over the considered duration. As the value of $a_\text{ini}$ increases, the VC increases considerably for all time instants.} \label{fig:EffectVC}
\end{figure}
We show in Fig.~\ref{fig:EffectVC} how the VC to the SFC changes over time for different values of $a_\text{ini}$. From this figure, we first note that the VC to the SFC increases during the late evening when there is a lack of solar energy due to lower intensity of solar irradiance. And, when the generation of solar energy is significantly high, i.e., around noon, the VC cost reaches a lower value and does not change significantly over time. This is due to the fact that a higher amount of solar generation eventually lets the SFC meet its demand from its own generation (case-2 and case-3) without any dependence on the grid. As a consequence, the VC reduces according to the proposed design. Whereas, the SFC needs to buy a significant amount of energy from the grid when there is not enough generation to meet the SFC's requirement (case-1). This consequently increases the SFC's estimate of cost in the next time slot, and hence the VC to the SFC increases considerably after time slot $20$ as shown in Fig.~\ref{fig:EffectVC}. We further note from Fig.~\ref{fig:EffectVC} that as the value of $a_\text{ini}$ increases the VC increases over all the time slots. Essentially, as explained in Section~\ref{subsection:cost function of SFC}, a higher $a_\text{ini}$ refers to a larger community size that requires more energy. Hence, the estimate of the cost across various time slots becomes larger compared to scenarios when $a_\text{ini}$ is small. As a consequence, VC attains a higher value for higher $a_\text{ini}$.

\subsection{Choice of ESD capacity}
\begin{figure}[t!]
\centering
\includegraphics[width=\columnwidth]{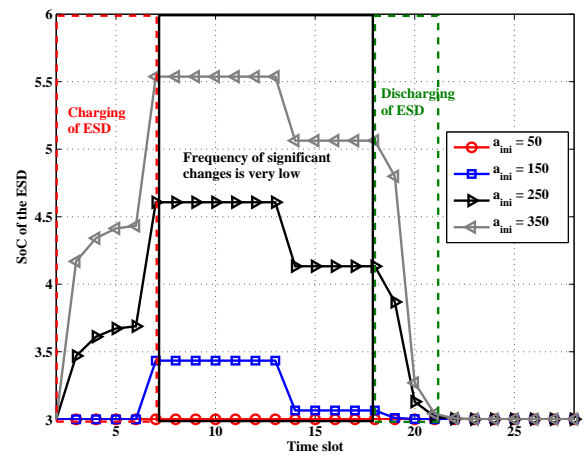}
\caption{Demonstration of the variation of SoC of SFC's ESD at different times of the day for different values of $a_\text{ini}$.} \label{fig:Battery-SoC}
\end{figure}

In Fig.~\ref{fig:Battery-SoC}, we demonstrate the charging and discharging behavior of the SFC's ESD for different values of $a_\text{ini}$, which would further provide some insights into the choice of the ESD's capacity. In the considered energy management scheme, we propose to use a cost factor in each charging and discharging cycle of the SFC to prevent excess charging and discharging of the SFC's ESD  during the energy management period. This enables the SFC to use a relatively lower amount of its capacity as demonstrated in this example. For instance, the SFC only uses up to $3.5$ kWh of ESD space for charging and discharging when $a_\text{ini} = 150$, which increases to $4.6$ and $5.5$ kWh for a $a_\text{ini}$ of $250$ and $350$ respectively. Interestingly, due to the possibility of increased average cost over the whole time duration, the SFC does not use its battery for $a_\text{ini} = 50$. Thus, for a fixed parameter $a_\text{ini}$, the proposed scheme enables the SFC to choose a suitable ESD capacity for the considered system, which can reduce the capital cost for the SFC by suggesting the set-up of a smaller-sized ESD.

Further from Fig.~\ref{fig:Battery-SoC}, the charging of the ESD mainly takes place in the morning when the PV generation is moderately high and the SFC discharges its ESD during the afternoon. Interestingly, no charging of the ESD is observed at noon even though the solar generation is significantly higher. This is due to the way the scheme is designed such that the cost factor $\alpha_b$ incurs a cost to the SFC whenever there is a charging or discharging of the ESD. Therefore, selling the surplus solar energy to households and the grid at noon enables the SFC to make more revenue instead of incurring a cost to it by charging the ESD. This strategy leads to a lower average cost to the SFC when considering the total time duration.

Furthermore, unlike most management schemes with energy storage devices where future information on price is available (e.g., see the management schemes surveyed in~\cite{Fang-J-CST:2012}) no charging of the ESD is observed in the late evening for the proposed technique. This is mainly due to two reasons: 1)  to increase the usage of renewable energy as explained in Section~\ref{sec:introduction}, we consider that the SFC only charges its ESD when the there is excess solar energy available and does not charge it with energy from the grid, and 2) it is assumed that no future information is considered available to the SFC. Therefore, the SFC decides on its energy management based on the scenarios available at each time slot, which prevents the SFC only from charging its ESD at night for future use.  However, it would be an interesting extension of the proposed work to include future information and observe the charging and discharging behavior of the ESD at different times.
\subsection{Impact of number of solar panels}
\begin{figure}[t!]
\centering
\includegraphics[width=\columnwidth]{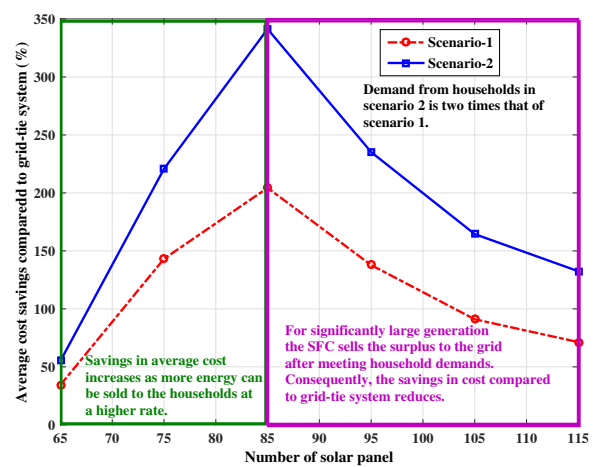}
\caption{Demonstration of the effect of change of solar panels on the percentage average cost savings to the SFC for the proposed scheme compared to grid-tie solar system~\cite{Nan-Deng:2011}.}
\label{fig:EffectSolarPanel}
\end{figure}

\begin{table*}[t!]
\centering
\caption{Demonstration of average cost savings to the SFC for the proposed scheme compared to a grid-tie system. The percentage improvements in terms of cost savings are shown in brackets.}
\includegraphics[width=0.8\textwidth]{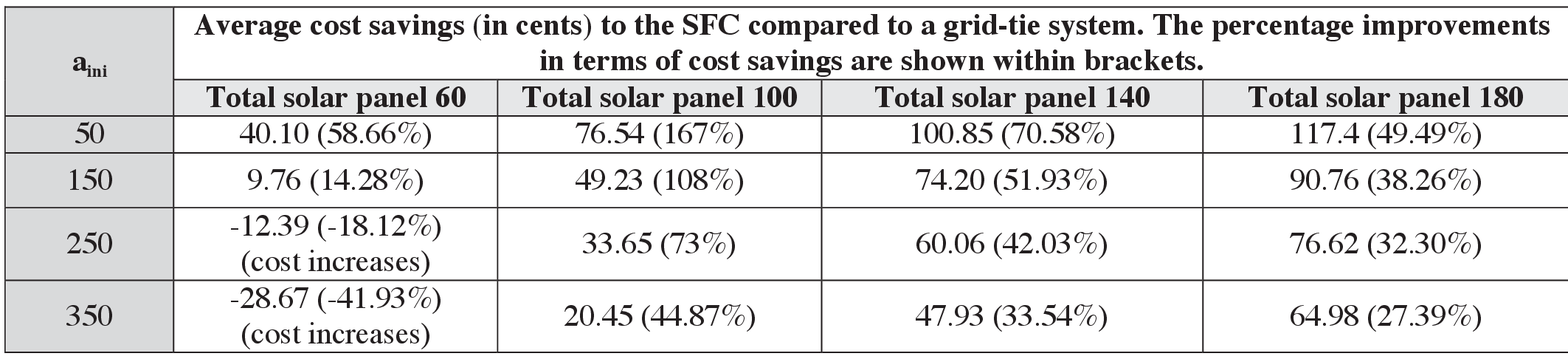}
\label{table:1}
\end{table*}

To compare the relative performance of the proposed scheme, we choose the grid-tie solar system~\cite{Nan-Deng:2011} as a benchmark, and show how the proposed scheme performs in terms of average cost savings to the SFC compared to a grid-tie solar system as the number of solar panels in the system changes. Essentially in a grid-tie system, an energy entity with renewables, such as the SFC in this paper, sells back its surplus energy, if there is any after meeting its own demand, to the grid. And if the generated energy is not enough to meet its requirements, the SFC may buy the deficient amount from the grid.

Now, to observe the performance improvement, we increase the number of solar panels of the SFC from $65$ to $115$ and demonstrate the percentage cost savings\footnote{Calculated using:$\frac{\text{cost for grid-tie system}-\text{cost for proposed scheme}}{\text{cost for grid-tie system}}\times 100$.} to the SFC for the  proposed scheme in comparison with the grid-tie system in Fig.~\ref{fig:EffectSolarPanel}. We consider two scenarios based on the household demands and consider that the household demand per time slot in scenario $2$ is twice that of scenario $1$.

In Fig.~\ref{fig:EffectSolarPanel}, first we note that the percentage cost savings for the SFC, in both scenarios, increases as the number of solar panels in the system increases from $65$ to $85$. Indeed, as the number of solar panels increases, the larger surplus enables the SFC to sell more to the households at a higher price compared to selling to the grid, which consequently increases the revenue for the SFC in the proposed scheme. Therefore, the percentage average cost savings with respect to the grid-tie system increases.

However, interestingly, as the number of solar panels increases from $90$ to $115$, the percentage of cost savings eventually decreases. This is because, as the scheme is designed, once the SFC fulfils both its own requirement and the requirements of the households, it sells the excess energy to the grid with a price similar to that of a grid-tie system. Hence, once the generation is significantly high (due to a large number of solar panels), most of the generated energy is sold back to the grid by the SFC. Therefore, the percentage improvement in terms of cost savings eventually decreases as the revenue from selling the energy to the grid is similar for both the proposed and grid-tie system.

Furthermore, as the households' demand in each time slots increases by a factor of two, the percentage cost savings increases (i.e., shifts upwards for scenario $2$ as can be seen from Fig.~\ref{fig:EffectSolarPanel}). Essentially, more household demand enables the SFC to sell more to the households, which increases its revenue and subsequently increases the average cost savings. Nonetheless, when the generation becomes significantly high in scenario 2, the cost savings reduce in a similar manner to scenario 1 for the same reason as for scenario 1.

\subsection{Impact of $a_\text{ini}$ on average cost savings}
Finally in Table~\ref{table:1}, we show how the average cost savings (in cents) to the SFC for the proposed scheme compared to a grid-tie system are affected for different choices of $a_\text{ini}$. The negative sign in the table implies that the cost for the proposed scheme is more than the grid-tie system. Now from Table~\ref{table:1}, as the value of $a_\text{ini}$ increases in a system, average cost savings to the SFC compared to the grid-tie system decreases for a particular number of solar panels. The main reason for this decrement can be explained from Fig.~\ref{fig:EffectVC}. According to Fig.~\ref{fig:EffectVC}, the VC to the SFC increases noticeably as $a_\text{ini}$ increases. That is, the SFC overestimates the cost in each time slot, which also contributes to the total cost to the SFC according to the design of the scheme. As a result, the cost to the SFC for the proposed scheme increases, which subsequently reduces the cost savings compared to a grid-tie system. For instance, for 60 solar panels, the average cost savings to the SFC from using the proposed scheme over a grid-tie system is 40.10 and 9.76 cents for $a_\text{ini}$ = $50$ and $150$ respectively, and the reduction of $30.34$ cents is due to the increment of VC to the SFC for a change of $a_\text{ini}$ by 100. And, as the value of $a_\text{ini}$ increases to $250$ and $350$, the proposed scheme shows a performance degradation compared to a grid-tie system.

However, if $a_\text{ini}$ is always set to a large value by the SFC, it can be interpreted that the community size is large and requires more energy. This subsequently means more solar panels need to be installed, if possible, in the system. Thus, the performance of the proposed scheme improves significantly for the system with higher demand and shows considerable cost savings when compared to a grid-tie system. For instance, the proposed scheme outperforms the grid-tie system at large values of $a_\text{ini}$ of 250 and 350 when the number of solar panels in the system is large, i.e., $140$ and $180$ respectively. Therefore, critical for the adaptation of the proposed scheme is the choice of $a_\text{ini}$ in accordance with the community size in order to capture a better cost-benefit tradeoff for the considered system. Further in terms of percentage improvement of the cost savings with respect to the number of solar panels, we note that the performance is similar to that in Fig.~\ref{fig:EffectSolarPanel}.

\section{Conclusion}
\label{sec:conclusion}
In this paper, we have proposed an energy management scheme for a shared facility controller (SFC), which is responsible for maintaining the shared facilities in a smart community and is also connected to the grid and households. A suitable system model has been proposed to enable multi-direction flow of electricity from the SFC's solar panels so as to minimize the operational cost to the SFC in each time slot of a considered duration. Considering the fact that the generation of energy from the SFC's solar panels and the requirement of energy for shared facilities are both intermittent, we have divided the energy management problem into three categories. In each category, the requirement of the SFC has been given a priority and the management scheme has been designed such that the SFC may also sell its excess electricity, if there is any, to other energy entities such as households and the grid. We have proposed the concept of a virtual cost (VC) and analyzed how the VC affects the decision making process of the SFC for three different cases. An algorithm has been proposed for the SFC to decide on the optimal charging and discharging amount of its ESD and on the trading of energy with different entities in real time in order to reach the optimal solution. Numerical studies have been provided to show the beneficial properties of the proposed scheme.

\end{document}